\documentclass[a4paper, 12pt]{scrartcl}
\usepackage[marginparwidth=35pt]{geometry}

\usepackage{lmodern}
\usepackage[english]{babel}
\usepackage[T1]{fontenc}
\usepackage[utf8]{inputenc}

\usepackage{graphicx}
\usepackage{float}

\usepackage{amsmath}
\usepackage{amsfonts}
\usepackage{amssymb}
\usepackage{amsthm}

\usepackage{caption}

\usepackage{bm}
\usepackage{bbold}

\usepackage{csquotes}

\usepackage{xspace}
\usepackage{xcolor}

\newcommand{\writeparameters}{${U_\uparrow = -2.25\text{ eV}}$, ${U_\downarrow = -0.1\text{ eV}}$, ${U_c=-3.}$, ${U_b = 0.77\text{ eV}}$, ${L=0.7\text{ nm}}$, ${L_\text{FL}=1\text{ nm}}$  and ${m_e^*/m_e = 0.8}$}

\title{Efficient solution strategy to couple micromagnetic simulations with ballistic transport in magnetic tunnel junctions}
\author{Peter Flauger$^1$, Claas Abert$^{1,2}$, Dieter Suess$^{1,2}$}

\begin{document}

\maketitle

\textsuperscript{1} Faculty of Physics, University of Vienna, Austria

\textsuperscript{2} University of Vienna Research Platform MMM Mathematics - Magnetism -  Materials, University of Vienna, Austria

\section*{Abstract}
We present a computationally efficient strategy that allows to simulate magnetization switching driven by spin-transfer torque in magnetic tunnel junctions within a micromagnetic  model coupled with a matrix-based non-equilibrium Green's function algorithm. Exemplary simulation for a realistic set of parameters are carried out and show switching times below $4\text{ ns}$ for voltages above $300\text{ mV}$ or around $2\times 10^{10}\text{Am}^{-2}$ for the ${\text{P}\rightarrow\text{AP}}$ (parallel to anti-parallel) direction. For ${\text{AP}\rightarrow\text{P}}$ switching, a trend-reversal in the switching time is seen i.e. the time for magnetization reversal first decreases with increasing bias voltage but then starts to rise again.

\section{Introduction}
A magnetic tunnel junction (MTJ) consists of two ferromagnetic layers that are connected by an insulating layer like MgO. The resistivity of such a stack depends on the angle between the two ferromagnet's magnetization directions~\cite{Julliere_1975_PhysLettA, Miyazaki_1995_JMMM}. This effect is called tunnel magnetoresistance (TMR) effect. It was further demonstrated that the magnetization direction of a magnetic layer in an MTJ can be manipulated by spin polarized currents \cite{Huai_2004_ApplPhysLett} due to a mechanism known as spin-transfer torque (STT) \cite{Slonczewski_1989_PhysRevB, Berger_1996_PhysRevB, Slonczewski_2007_JMMM}. The combination of the two effects allows for a persistent memory technology called magnetoresistive random access memory (STT-MRAM) that is of technological interest \cite{Huai_2008_AAPPS,Bhatti_2017_MaterToday}.\\
Building on the pioneering work of Caroli et al. \cite{Caroli_1971_JPhysC}, theoretical attempts to describe the behavior of STT in MTJs analytically have been made by \cite{Theodonis_2006_PhysRevLett, Manchon_2008_JPhysCondensMatter, Kalitsov_2009_PhysRevB, Chshiev_2015_PhysRevB} and describe the torques arising from a free electron model within the framework of the Keldysh formalism and the WKB approximation in great detail. On the basis of reference~\cite{Theodonis_2006_PhysRevLett}, Kubota et al. \cite{Kubota_2007_NPhys} showed a qualitative agreement to their experimental findings and their work was used in turn by Datta et al. in reference~\cite{Datta_2011_IEDM} as basis for the parameters in their computational work.\\\\
Although it is a standard technique in micromagnetism to couple magnetization dynamics to spin-drift-diffusion calculations to describe a vast variety of GMR-stack based devices~\cite{Strelkov_2011_PhysRevB, Abert_2016_SciRep}, the same does not hold true for devices based on MTJs. Due to the ballistic nature of the transport in MTJs, a local solution is insufficient and coupling of magnetization dynamics with matrix-based approaches to solve for the wavefunctions within the insulating region as presented e.g. in reference~\cite{Salahuddin_2006_ApplPhysLett} are generally very expensive from a computational point of view. 
% referee response
The computational cost of this approach arises mainly due to the involved matrix inversion. Analytical solutions like the ones presented in reference \cite{Chshiev_2015_PhysRevB} can be used to calculate the transport properties instead, with comparably little computational load. However, analytical solutions are derived for predetermined stack configurations and are thus limited to them. Similar problems are faced when using heuristic models in combination with parameters extracted from experiments. This approach is further limited by data availability: while there are a lot of studies on the electrical transport properties through a variety on MTJ stack configurations, the amount of data on voltage dependent angular momentum transport like in references~\cite{Kubota_2007_NPhys,Sankey_2008_NPhys} is sparse.\\
The general versatility of numerical solutions remains their biggest draw. Thus, there is a demand for efficient ways to introduce numerically obtained transport properties into magnetization dynamics simulations.
\\After giving an introduction to both magnetization dynamics and the non-equilibrium Green's function formalism, we will present a solution strategy that results in highly performant algorithms for fixed-voltage and fixed-current simulations. The paper is organized as follows: section~\ref{sec:methods} outlines how to introduce torque to the Landau-Lifshitz-Gilbert equation and how this torque can be obtained from the non-equilibrium Green's function formalism. In section~\ref{sec:modeling}, a solution strategy based on the torque's properties is proposed and simulated switching behavior of an STT-MRAM cell is presented and discussed in section~\ref{sec:results}. 

\section{Methods}
\label{sec:methods}
Within the framework of micromagnetism, the magnetization dynamics in an MTJ can be described by the Landau-Lifshitz-Gilbert equation

\begin{equation}
\frac{\partial\bm{m}}{\partial t} = -\gamma\bm{m}\times\bm{H}^\text{eff}+\alpha\bm{m}\times\frac{\partial\bm{m}}{\partial t}+\bm{T}
\label{eq:LLG}
\end{equation}
where $\gamma$ denotes the gyromagnetic ratio, $\bm{m}$ is the local magnetization unit vector field, $\alpha$ the Gilbert damping parameter and $\bm{H}^\text{eff}$ the effective field acting on the local magnetization. $\bm{T}$ is the torque due to the magnetic moments that are transported through the barrier region of the MTJ. This torque can be obtained from non-equilibrium Green's function (NEGF) calculations \cite{Datta_2011_IEDM, Kalitsov_2009_PhysRevB} as described in the following: Assuming that the local magnetic moment can be attributed to localized d-electrons while charge and spin are transported by s-electrons, the properties of the latter can be modeled by the device Hamiltonian

\begin{equation}
\hat{H}(\epsilon_\perp,x)=\left[\frac{\hat{p}^2}{2m^*_e}+u(x)+\epsilon_\perp\right]I_2+J_\text{sd}(x)\bm{m}(x)\cdot\bm{\sigma}.
\end{equation}
Here, $m_e^*$ is the effective mass of the s-electrons, $u(x)$ is the local potential energy, $J_{sd}$ the exchange coupling energy between the delocalized and localized electrons and $\epsilon_\perp$ the energy in transverse momentum component i.e. perpendicular to the current direction and parallel to the interfaces between materials, so that the kinetic energy can be expressed by $\epsilon-u(x) = \epsilon_\perp+\epsilon_\parallel$. This Hamiltonian is then brought into its discretized form in order to solve the Schrödinger equation truncated to the device region \cite{Gaury_2014_PhysRep} 

\begin{equation}
\left[\epsilon-\hat{H}(\epsilon_\perp,x)-\hat{\Sigma}^R(\epsilon_\parallel,x)\right]\psi(\epsilon,\epsilon_\perp,x)\Big |_{x\in [x_0,x_{N-1}]} = \hat{S}(\epsilon_\parallel,x)
\end{equation}
with the open boundary conditions $\hat{\Sigma}^R(\epsilon_\parallel,x)$ and a source term of the form

\begin{equation}
\hat{S}(\epsilon_\parallel,x) = \sqrt{v_L(\epsilon_\parallel)}\delta(x-x_L) + \sqrt{v_R(\epsilon_\parallel)}\delta(x-x_R).
\end{equation}
The retarded Green's function $\hat{G}^R(\epsilon,\epsilon_\perp,x,x')$ is obtained by solving equation~(\ref{eq:RetardedGreensFunction})

\begin{equation}
\left[\epsilon-\hat{H}(\epsilon_\perp,x)-\hat{\Sigma}^R(x,\epsilon_\parallel)\right]\hat{G}^R(\epsilon,\epsilon_\perp,x,x') = \delta(x-x')
\label{eq:RetardedGreensFunction}
\end{equation}
and thus

\begin{equation}
\hat{G}^R(\epsilon,\epsilon_\perp,x,x')*\hat{S}(\epsilon_\parallel,x') = \psi(\epsilon,\epsilon_\perp,x)
\end{equation}
holds true. Within the tight-binding approximation \cite{Yanik_2007_PhysRevB}, the discretized Hamiltonian with site-to-site hopping energy 

\begin{equation}
t=\frac{\hbar^2}{2m^*_e\Delta x^2}
\end{equation}
reads 

\begin{equation}
H_{ij}(\epsilon_\perp) = \begin{cases}
\left(u_i+2t+\epsilon_\perp\right)I_2 + J_\text{sd}\bm{m}_i\cdot\bm{\sigma} & i=j \\
-tI_2 & j=i\pm 1 \\
0_{2,2} & \text{else}
\end{cases}
\end{equation}
The open boundary conditions

\begin{equation}
\Sigma^R_L(\epsilon_\parallel) = -t R_L \left[\begin{array}{cc} 
\exp{\left(ik_L^\uparrow\Delta x\right)} & 0 \\ 
0 & \exp{\left(ik_L^\downarrow\Delta x\right)} \\ 
\end{array}\right] R_L^\dagger
\end{equation}
and

\begin{equation}
\Sigma^R_R(\epsilon_\parallel) = -t R_R \left[\begin{array}{cc} 
\exp{\left(ik_R^\uparrow\Delta x\right)} & 0 \\ 
0 & \exp{\left(ik_R^\downarrow\Delta x\right)} \\ 
\end{array}\right] R_R^\dagger
\end{equation}
with the rotation matrices $R_L$ and $R_R$ from the z-axis to the local spin-quantization axis i.e. the local magnetization direction in the left and right lead respectively are included in the self-energy matrix:

\begin{equation}
\Sigma^R_{ij}(\epsilon_\parallel) = \begin{cases}
\Sigma_L^R(\epsilon_\parallel) & i=j=0 \\
\Sigma_R^R(\epsilon_\parallel) & j=i=N-1 \\
0_{2,2} & \text{else}
\end{cases}
\end{equation}
After calculating the retarded Green's function by matrix inversion via equation~(\ref{eq:Inversion})

\begin{equation}
G^R(\epsilon,\epsilon_\perp) = [\epsilon I_{2N}-H(\epsilon_\perp)-\Sigma^R(\epsilon_\parallel)]^{-1}
\label{eq:Inversion}
\end{equation}
the kinetic equation~(\ref{eq:Kinetic}) with the in-scattering function (\ref{eq:SigIn}) is solved.
\begin{equation}
\Sigma_{ij}^\text{in}(\epsilon,\epsilon_\parallel) = \begin{cases}
i f_L(\epsilon)\left(\Sigma_L^R(\epsilon_\parallel)-\Sigma_L^{R\dagger}(\epsilon_\parallel)\right) & i=j=0 \\
i f_R(\epsilon)\left(\Sigma_R^R(\epsilon_\parallel)-\Sigma_R^{R\dagger}(\epsilon_\parallel)\right) & j=i=N-1 \\
0_{2,2} & \text{else}
\end{cases}
\label{eq:SigIn}
\end{equation}

\begin{equation}
G^n(\epsilon) = \int G^R(\epsilon,\epsilon_\perp)\Sigma^{in}(\epsilon,\epsilon_\perp)G^A(\epsilon,\epsilon_\perp) d\epsilon_\perp
\label{eq:Kinetic}
\end{equation}
The two functions $f_L(\epsilon)$ and $f_R(\epsilon)$ are the occupation functions (Fermi-Dirac distributions) for the left and right lead, respectively. This procedure yields the non-equilibrium Green's function (\ref{eq:Gn}), which is proportional to the local density matrix. A more detailed discussion of this method can be found in reference~\cite{Datta_2000_SuperlatticeMicrost}, although not for its spin-polarized form. 

\begin{equation}
G^n_{i,j}(\epsilon) = 2\pi\Delta x\left[\psi(\epsilon,x_i)^\dagger\psi(\epsilon,x'_j)\right]
\label{eq:Gn}
\end{equation}
The local spin accumulation

\begin{equation}
s_k(x_i) = \frac{\mu_B}{2\pi\Delta x}\int \operatorname{Tr_\sigma}\left[ G_{ii}^n(\epsilon) \sigma_k \right] d\epsilon
\label{eq:s_from_Gn}
\end{equation}
is the origin of the torque acting on the magnetization due to the coupling between s- and d-electrons \cite{Kalitsov_2006_JApplPhys, Chshiev_2015_PhysRevB}, as described by 

\begin{equation}
\bm{T} = \frac{\partial \bm{m}}{\partial t}\Big|_\text{sd-coupling} = -\frac{J_{sd}}{\hbar M_s}\bm{m}\times\bm{s}.
\label{eq:torque}
\end{equation}
It is a common technique \cite{Slonczewski_1989_PhysRevB} to separate the torque into its dampinglike and fieldlike contributions, $T_\text{dl}$ and $T_\text{fl}$ respectively. The base vectors of the torque components are obtained as stated in (\ref{eq:torque_fieldlike}) and (\ref{eq:torque_dampinglike}) from the local magnetization and the magnetization of the source region $\bm{p}$ -- that is the magnetization at the left interface for the right lead and vice versa.

\begin{align}
&\bm{T} = \bm{T}_\text{dl} + \bm{T}_\text{fl} \\
&\bm{T}_\text{fl} = \tau_\text{fl}\,\bm{m}\times\bm{p} \label{eq:torque_fieldlike}\\
&\bm{T}_\text{dl} = \tau_\text{dl}\,\bm{m}\times \left( \bm{m}\times\bm{p} \right) \label{eq:torque_dampinglike}
\end{align}

\section{Modeling}
\label{sec:modeling}
\begin{figure}[H]
\includegraphics[width=\textwidth]{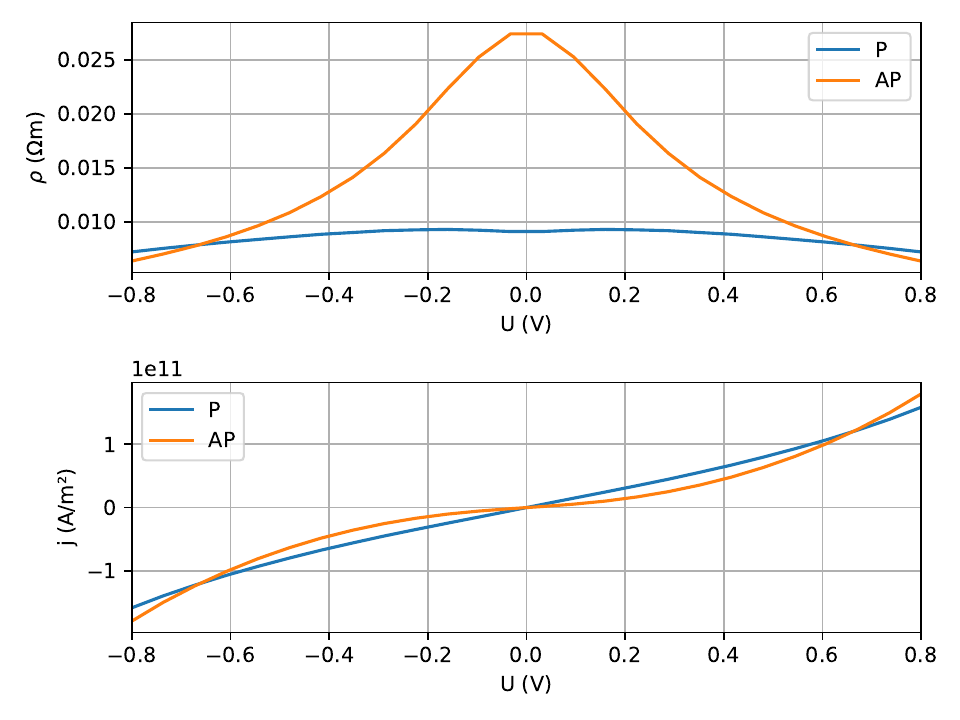}
\caption{Voltage dependence of the resistivity and current density for the parallel (P) and anti-parallel (AP) magnetization configuration of a symmetric MTJ with two semi-infinite ferromagnetic leads. The parameters are ${U_\uparrow = -2.25\text{ eV}}$, ${U_\downarrow = -0.1\text{ eV}}$, ${U_b = 0.77\text{ eV}}$, ${L=0.7\text{ nm}}$ and ${m_e^*/m_e = 0.8}$.}
\label{fig:rho_voltage_dependence}
\end{figure}

\begin{figure}[H]
\includegraphics[trim=100 0 0 0, width=\textwidth]{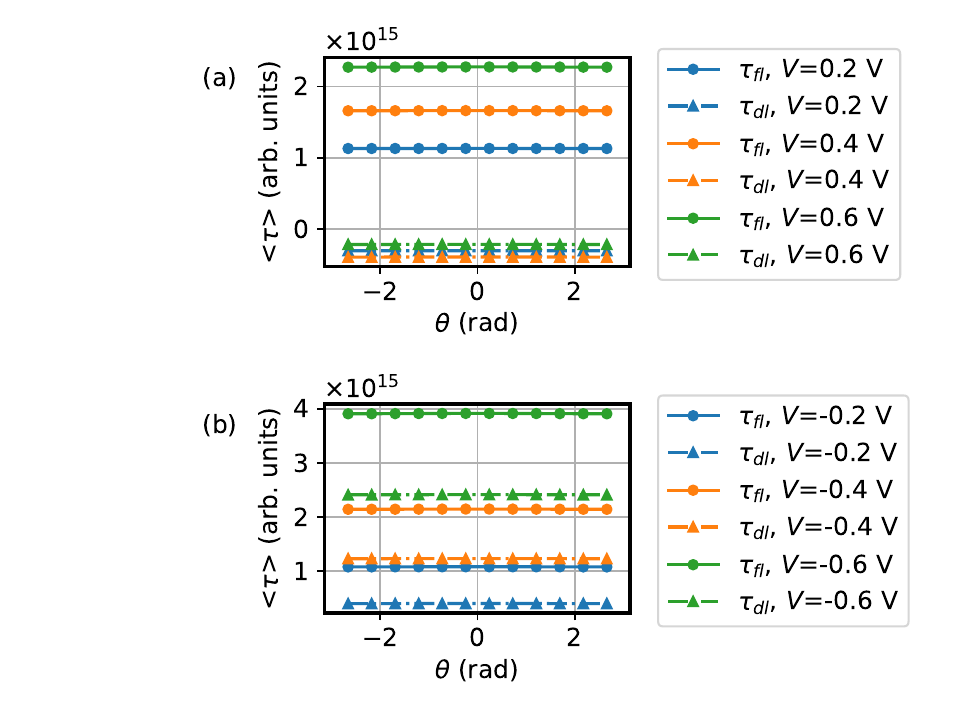}
\caption{Angular independence of the average torque components (circles mark the fieldlike torque component, triangles mark the dampinglike torque component) for different voltages. (a) and (b) illustrate the torques acting on the free-layer for positive voltages and negative voltages, respectively.}
\label{fig:angular_dependence_tau}
\end{figure}

For all calculations in this work, the  MTJs are characterized by the spin-dependent potentials $U_\uparrow$ and $U_\downarrow$ inside the magnetic material, the barrier height $U_b$ within the insulating layer and the spin-independent potential $U_c$ within the conducting lead attached to one of the magnetic layers, such that 

\begin{align}
u(z) &= \begin{cases}
U_\text{FM}-eV/2 & z < z_L \\
U_b + \frac{eV\left(z-z_L\right)}{z_R-z_L}-eV/2  & z_L \le z \le z_R \\
U_\text{FM}+eV/2 & z_R < z \le z_R + L_\text{FL} \\
U_c  + eV/2 & z > z_R + L_\text{FL}
\end{cases} \label{eq:potential function} \\
U_\text{FM} &= \left( U_\downarrow + U_\uparrow \right)/2 \\
J_\text{sd} &= \left( U_\downarrow - U_\uparrow \right)/2.
\end{align}
MTJs that can be described by such a function could be seen to be symmetric with respect to their parameters. The modeling of the device via (\ref{eq:potential function}) assumes a semi-infinite magnetic layer on one side of the structure and a semi-infinite non-magnetic layer on the other side. In this sense, the junction is asymmetric. The thickness of the insulating layer is $L=z_R-z_L$ and the thickness of the finite-sized magnetic layer (that will act as free-layer in the micromagnetic simulations) is $L_\text{FL}$. The applied voltage $V$ introduces a linear change in the barrier potential and shifts the Fermi levels in the leads. When doing the integration of equation~(\ref{eq:s_from_Gn}), it is sufficient to consider only energies that lie between the two Fermi levels since no other state can contribute to the current flow and the spin accumulation.\\
If not stated otherwise, all results are obtained for \writeparameters , mostly in accordance with reference~\cite{Datta_2011_IEDM}.\\\\
As depicted in figure~\ref{fig:rho_voltage_dependence}, the resistivity and the TMR ratio of a tunnel junction depend on the applied voltage. The TMR ratio of an MTJ can be measured either with a fixed current or a fixed voltage. For the selected set of parameters, the two torque components $\tau_\text{fl}$ and $\tau_\text{dl}$ do not depend on the angle $\theta$ between the reference and free-layer magnetization directions for fixed voltages \cite{Sankey_2008_NPhys}. This can be seen in figure~\ref{fig:angular_dependence_tau}. Therefore, it is possible to pre-compute $\tau_\text{fl}$ and $\tau_\text{dl}$ for micromagnetic simulations with a fixed voltage, such that the torque can be obtained from equations~(\ref{eq:torque_fieldlike}) and (\ref{eq:torque_dampinglike}) for every timestep of the LLG integration. Since the matrix inversion in (\ref{eq:Inversion}) is computationally expensive, the performance of the whole algorithm will greatly benefit from this procedure.\\\\
The full computation strategy for fixed voltages is then as follows: For the NEGF calculations a finite differences 1D-mesh is set up along the direction perpendicular to the interfaces that truncates the semi-infinite leads problem to the region of interest. For the micromagnetic simulations, either a finite differences or finite elements mesh for the whole device is generated. This mesh might also include non-magnetic leads or other extensions to the MTJ-stack and can be discretized much coarser than the mesh used for the transport problem.\\
Next, equation~(\ref{eq:s_from_Gn}) is solved for $\theta=\angle\left(\bm{m}\left(z_L\right), \bm{m}\left(z_R\right)\right)=\pi /2$ and the resulting accumulation $\bm{s}(V,z)$ is projected onto ${-\bm{m}\times\left(\bm{m}\times\bm{p}\right)}$ and ${\bm{m}\times\bm{p}}$, yielding $s_\text{fl}$ and $s_\text{dl}$ such that 

\begin{equation}
	\bm{s}^\text{eff}(V,z)= -s_{\text{fl}}\bm{m}\times\left(\bm{m}\times\bm{p}\right) + s_{\text{dl}}\bm{m}\times\bm{p}
\end{equation}
results in the same torque as the original $s(V,z)$ when plugged into equation~(\ref{eq:torque}). Equation~(\ref{eq:torque}) also relates $s_\text{fl}$ and $s_\text{dl}$ with $\tau_\text{fl}$ and $\tau_\text{dl}$. Like their torque-counterparts, $s_\text{fl}$ and $s_\text{dl}$ do not change with $\theta$ and torque for every configuration of $\bm{m}$ and $\bm{p}$ can be calculated from them. Note that the component of $\bm{s}$ parallel to $\bm{m}$ cannot contribute to the exerted torque. 
\\After storing $s_\text{fl}$ and $s_\text{dl}$, the integration of the LLG can be started and for every timestep, the contributions to $\bm{H}^\text{eff}$ have to be computed. Contributions might be independent of the vector field $\bm{m}$ like e.g. an external field or might depend on the magnetization like e.g. the exchange field. As can be seen easily from equation~(\ref{eq:LLG}), the effective field contribution equivalent to $\bm{T}$ is just

\begin{equation}
\bm{H}^\text{Torque}=\frac{J_\text{sd}}{\hbar M_s \gamma}\bm{s}.
\end{equation}
Hence, in every region of the device $\bm{m}$ and $\bm{p}$ are read out from the respective interface values according to the rule 

\begin{equation}
\bm{p} = \begin{cases}
\bm{m}\left(z_R\right) & z<z_L \\
\bm{m}\left(z_L\right) & z>z_R \\
\end{cases}
\end{equation}
and the local effective field due to the torque is calculated after constructing $\bm{s}^\text{eff}$ from the resulting basis vectors and the stored values of $s_\text{fl}$ and $s_\text{dl}$. Once again, this reconstructed value of the spin accumulation has no component along the local magnetization direction and differs therefore from the actual value that would be obtained from equation~(\ref{eq:s_from_Gn}) for the current magnetization configuration. However, the resulting torques do not differ since no contribution of the effective field $\bm{H}^\text{eff}$ along $\bm{m}$ influences the dynamics yielded by the LLG.\\
In the case of fixed-current-experiments, the torque components depend on the angle $\theta$. Still, the strategy described above can be adapted by computing a look-up-table for different values of $\theta$ before simulating the magnetization dynamics, so that the torque components can then be interpolated from the stored values.\\\\
The descretization chosen for the transport problem will be at least one order of magnitude smaller than the descretization of the mesh for the magnetization dynamics problem and thus, the torque obtained from the former should be averaged to be used with the later. The average values of $\tau_\text{fl}$ and $\tau_\text{dl}$ are obtained over the length of the ferromagnetic free-layer $L_\text{FL}$ \cite{Manchon_2008_JPhysCondensMatter,Chshiev_2015_PhysRevB}. 

\begin{equation}
\left\langle \tau \right\rangle =  \frac{1}{L_\text{FL}} \int_{z_R}^{z_R+L_\text{FL}} \tau (z)dz
\label{eq:tau_mean}
\end{equation}
This method of averaging is justified by assuming that (i) the micromagnetic assumption is not violated i.e. the magnetization only changes slightly from mesh node to mesh node and (ii) $L_\text{FL}$ is shorter than the mean free path length for s-electrons in the ferromagnetic leads. It should be noted that while no scattering is taken into account, reflections of both interfaces of the free-layer are. 

\begin{figure}[H]
\includegraphics[width=\textwidth]{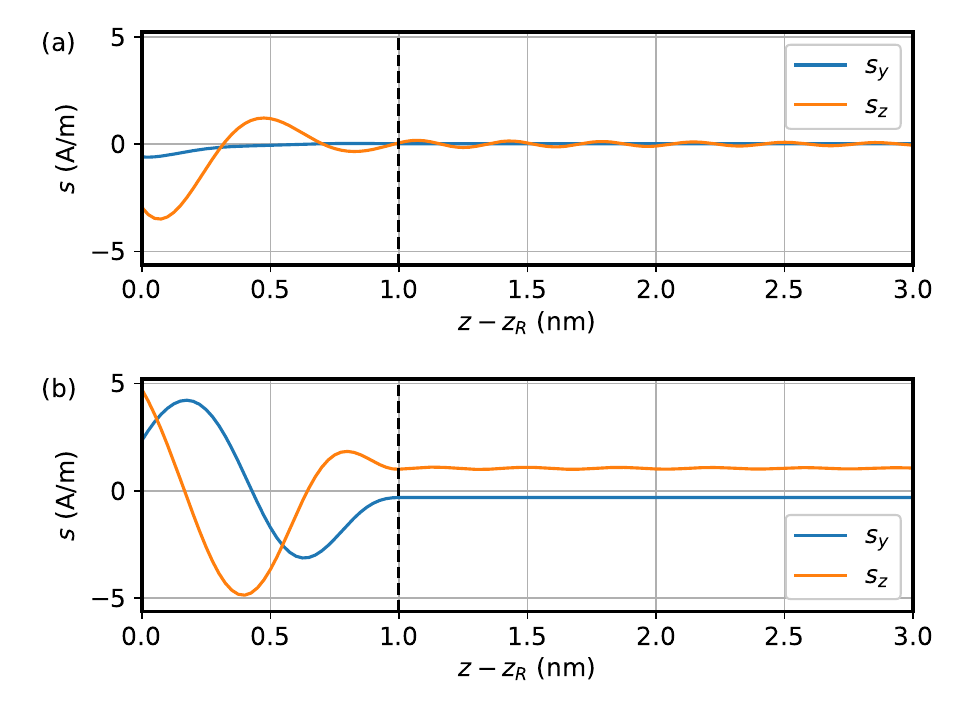}
\caption{Spatial dependence of the spin accumulation components orthogonal to the local magnetization direction ($z$-direction) in the free-layer for positive and negative voltages. (a) shows the spin accumulation components for $V=350\text{ mV}$ i.e. electrons are injected from the right, (b) for $V=-350\text{ mV}$ i.e. electrons are injected from the left. The length of the ferromagnetic lead is assumed to be ${L_\text{FM}=1\text{ nm}}$, indicated above by a dashed vertical lines.}
\label{fig:s_spatial_dependence}
\end{figure}

\section{Results and Discussion}
\label{sec:results}
\begin{figure}[H]
\includegraphics[width=\textwidth]{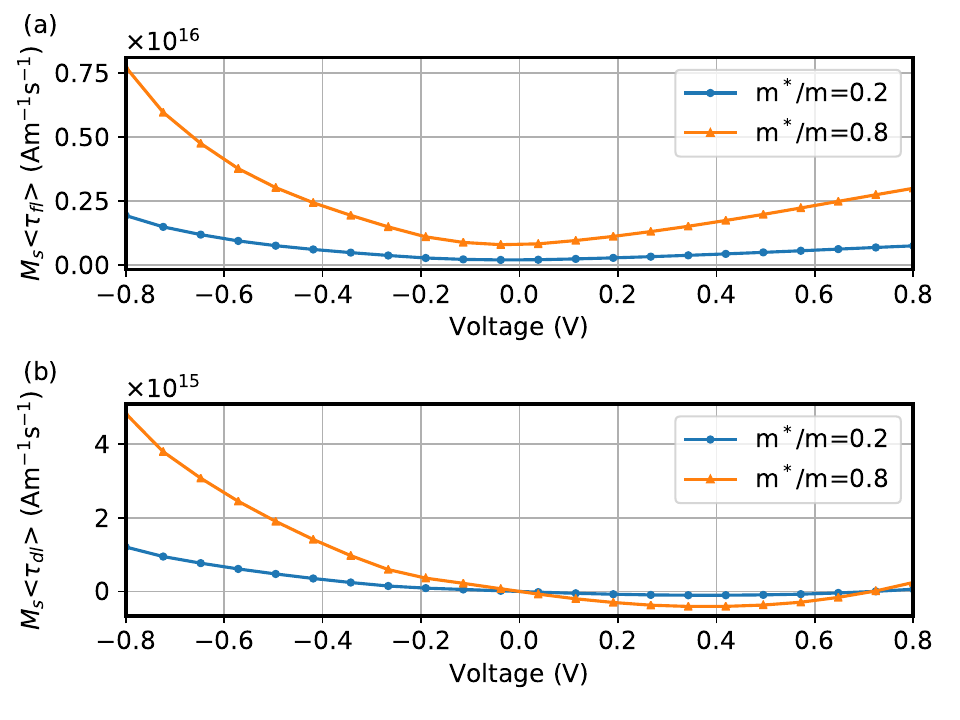}
\caption{Voltage dependence of the fieldlike (a) and dampinglike (b) torque components. While the fieldlike torque component stays positive weather the bias voltage is positive or negative, the dampinglike torque has a zero-crossing at zero bias. This enables STT-based switching. However, for voltages above $\approx 700\text{ mV}$, there is another sign-change in this component and bi-directional switching becomes impossible.}
\label{fig:voltage_dependence_tau}
\end{figure}

\begin{figure}[H]
\includegraphics[width=\textwidth]{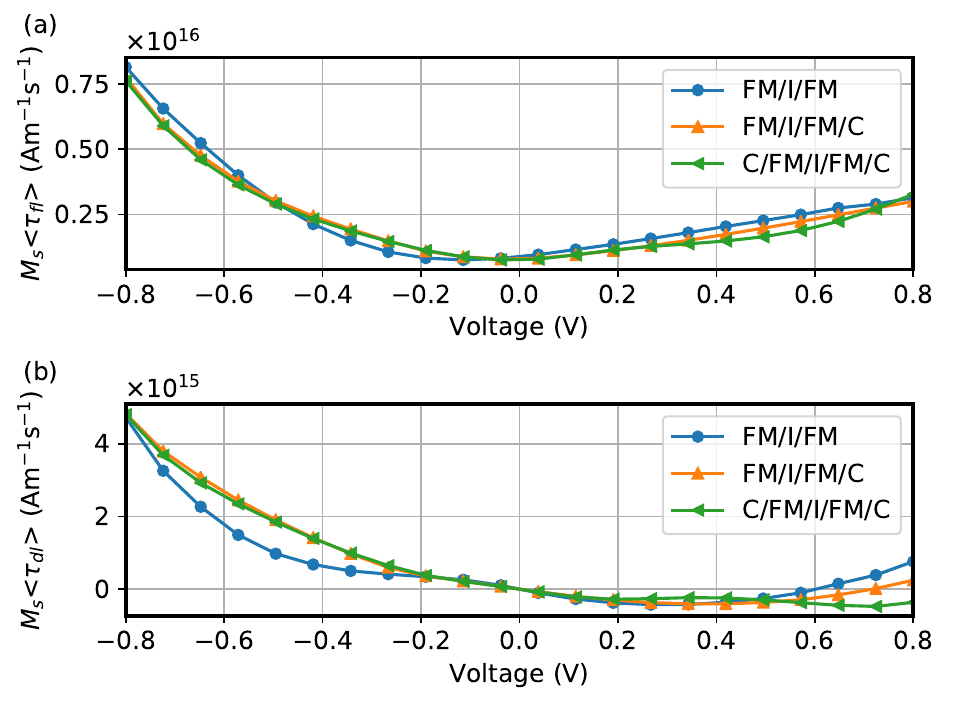}
\caption{Voltage dependence of the fieldlike (a) and dampinglike (b) torque components for different stacks. All mean values are obtained for ${L_\text{FM}=1\text{ nm}}$. The blue lines (circles) correspond to a symmetric stack with semi-infinite ferromagnetic leads. The orange lines (upward triangles) are the same as in figure~\ref{fig:voltage_dependence_tau} for $m_e^*/m_e=0.8$ and correspond to a stack with a semi-infinite reference-layer and a $1\text{ nm}$ free-layer followed by a semi-infinite non-magnetic conductor. The green lines (left triangles) correspond to a stack that has two finite-sized ferro-magnets of $1\text{ nm}$ thickness each attached to the insulator.}
\label{fig:voltage_dependence_tau_diff_stacks}
\end{figure}

A functional STT-based MRAM cell can be switched from the parallel configuration (P) to the anti-parallel configuration (AP) and vice versa by changing the polarity of the applied voltage accordingly. This form of bidirectional switching is only possible due to the sign-reversal of the dampinglike torque with respect to the current direction i.e. ${\left\langle\tau_\text{dl}\right\rangle(V>0)>0}$ and ${\left\langle\tau_\text{dl}\right\rangle(V<0)<0}$ or ${\left\langle\tau_\text{dl}\right\rangle(V>0)<0}$ and ${\left\langle\tau_\text{dl}\right\rangle(V<0)>0}$. The voltage dependence of the averaged values of $\tau_\text{fl}$ and $\tau_\text{dl}$ in the presented simulations can be inspected in figure~\ref{fig:voltage_dependence_tau}. One curious feature of the depicted behavior is a trend-reversal of the dampinglike torque for voltages above $\approx 0.35\text{ V}$. This indicates a decrease in the switching efficiency / switching time for increasing currents from a certain point on for the ${\text{PA}\rightarrow\text{P}}$ switching direction until switching becomes impossible as a result of the sign-reversal of the dampinglike torque for $V>0.7\text{ V}$. Meanwhile the fieldlike contribution seems to increase with the voltage, independent of its polarity. This observation is not a novel one and similar behavior can be seen and was noted in theoretical studies of torques in MTJs from analytical descriptions in~\cite{Theodonis_2006_PhysRevLett, Manchon_2008_JPhysCondensMatter, Kalitsov_2009_PhysRevB, Chshiev_2015_PhysRevB}. Comparison to measurements in~\cite{Kubota_2007_NPhys} confirmed this behavior, although its ramifications are generally not observed in experimentally obtained switching probabilities and critical currents measurements.\\\\
Figure~\ref{fig:voltage_dependence_tau_diff_stacks} shows how different models of the MTJ stack influence the behavior of the torque components. The first case (FM/I/FM) models the tunnel junction to consist of two semi-infinite ferromagnetic leads attached to an insulator. The second case (FM/I/FM/C) conforms to equation~(\ref{eq:potential function}) and and differs from the FM/I/FM case in that it includes reflections from the free-layer/conductor interface. The inclusion of this reflections yields larger absolute values of dampinglike torque. The last depicted case (C/FM/I/FM/C) assumes that the length of the reference-layer is also $1\text{ nm}$ and reflections from the reference-layer/conductor interface are also included. In all cases, the mean values are obtained from equation~(\ref{eq:tau_mean}) with ${L_\text{FM}=1\text{ nm}}$. The different structures of the stack are easily modeled by modifying equation~(\ref{eq:potential function}) accordingly. This flexibility is one of the key advantages of numerical solutions compared to analytical ones.

\begin{figure}[H]
\includegraphics[width=\textwidth]{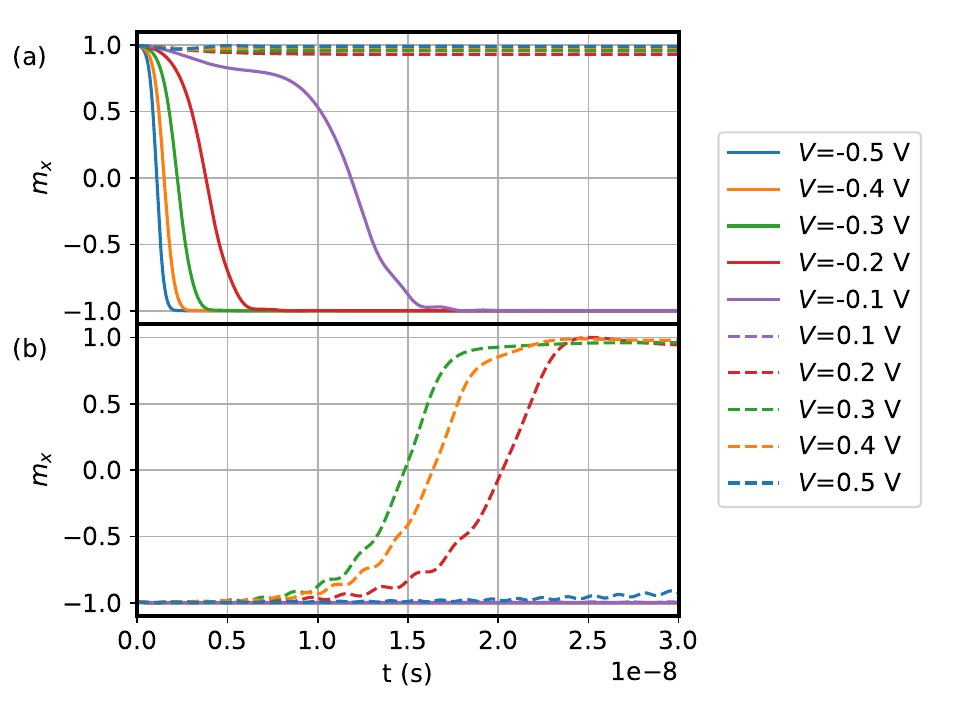}
\caption{Simulation of the magnetization dynamics in the coupled LLG+NEGF system for different values of the bias voltage $V$. Simulations in (a) are initialized with their magnetization of the free-layer almost parallel to the reference-layer magnetization i.e the positive $x$-direction. (b) shows simulations that are initialized with an anti-parallel configuration.}
\label{fig:switching}
\end{figure}

\begin{figure}[H]
\includegraphics[width=\textwidth]{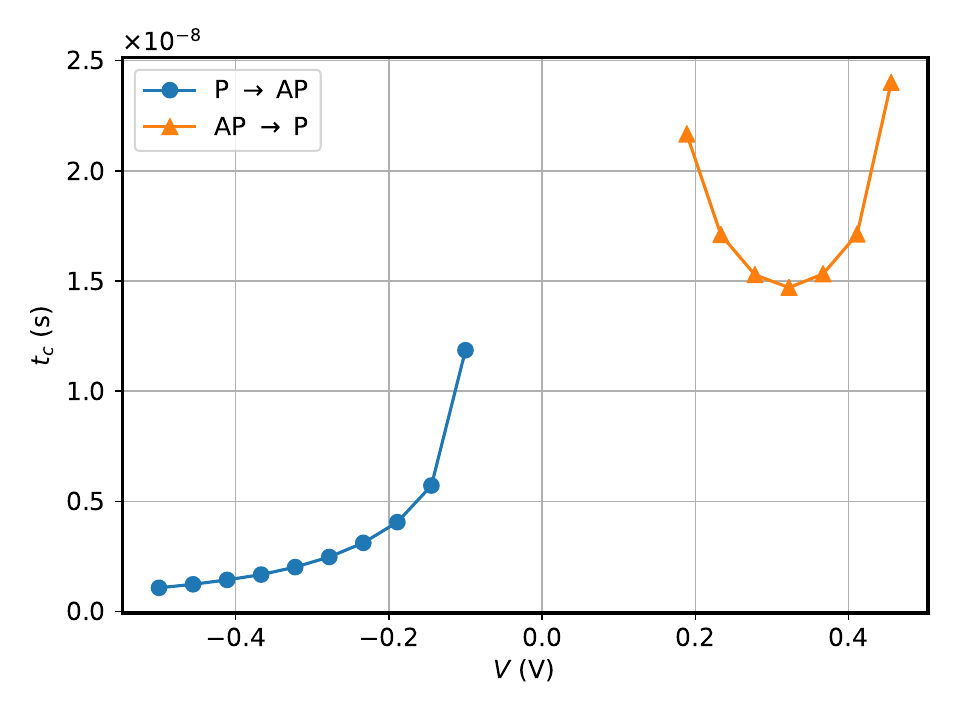}
\caption{Voltage-dependence of the critical pulse duration $t_c$ i.e. the minimum pulse duration required for deterministic switching.}
\label{fig:critical_current}
\end{figure}

The micromagnetic simulations are achieved by the python library magnum.fe \cite{Abert_2013_JMMM}. Integration of the modified LLG given by equation~(\ref{eq:LLG}) coupled with the spin accumulation from equation~(\ref{eq:s_from_Gn}) via equation~(\ref{eq:torque}) yields the switching behavior shown in figure~\ref{fig:switching}. In these micromagnetic simulations, the FM/I($0.7$)/FM($1$) structure is stacked along the $z$-direction. The values in the brackets indicate the layer thickness in nm. The lower ferromagnetic layer is the reference-layer ($z<z_L$), the upper layer is the free-layer ($z>z_R$). Note that the reference-layer exhibits no magnetization dynamics and its size in the micromagnetic simulations is thus arbitrary while it is modeled semi-infinite by equation (\ref{eq:potential function}) in the context of the NEGF calculations. The results therefore resemble the behavior of an MTJ with very thick reference-layer. The saturation magnetizations for the reference- and free-layer are $M_s=1.24/(4\pi)\times 10^{7}\text{ Am}^{-1}$ and $1/(4\pi)\times 10^{7}\text{ Am}^{-1}$, respectively. The uniaxial anisotropy of the reference-layer ($K_u=10^6\text{ Jm}^{-3}$) with the easy axis along the $x$-direction fixates the local magnetization. In the free-layer, the effective uniaxial anisotropy constant is set to be $K_u=4\times 10^{3} \text{ Jm}^{-3}$ and the anisotropy axis is tilted by $\approx 5^\circ$ out of the $x$-direction towards the $z$-direction. The anisotropy is labeled effective, since the demagnetization field is not taken into account explicitly but rather its effects are included in the anisotropy field term. In both ferromagnets, the exchange coupling is $A=2.8\times 10^{-11}\text{ Jm}^{-1}$ and the Gilbert damping parameter $\alpha$ is $0.08$.\\\\
The ${\text{P }\rightarrow\text{ AP}}$ switching shown in figure~\ref{fig:switching}(a) is enabled by negative voltages and requires less current flow than ${\text{AP }\rightarrow\text{ P}}$ switching, which is illustrated in figure~\ref{fig:switching}(b). The minimal writing-pulse duration $t_c$ for deterministic switching is plotted against the applied bias voltage in figure~\ref{fig:critical_current} and can be compared easily to figure \ref{fig:switching}. As already discussed in the context of figure~\ref{fig:voltage_dependence_tau}, the critical pulse duration $t_c$ for ${\text{AP }\rightarrow\text{ P}}$ switching does indeed increase for voltages above $\approx0.35\text{ V}$.

\begin{figure}[H]
\includegraphics[width=\textwidth]{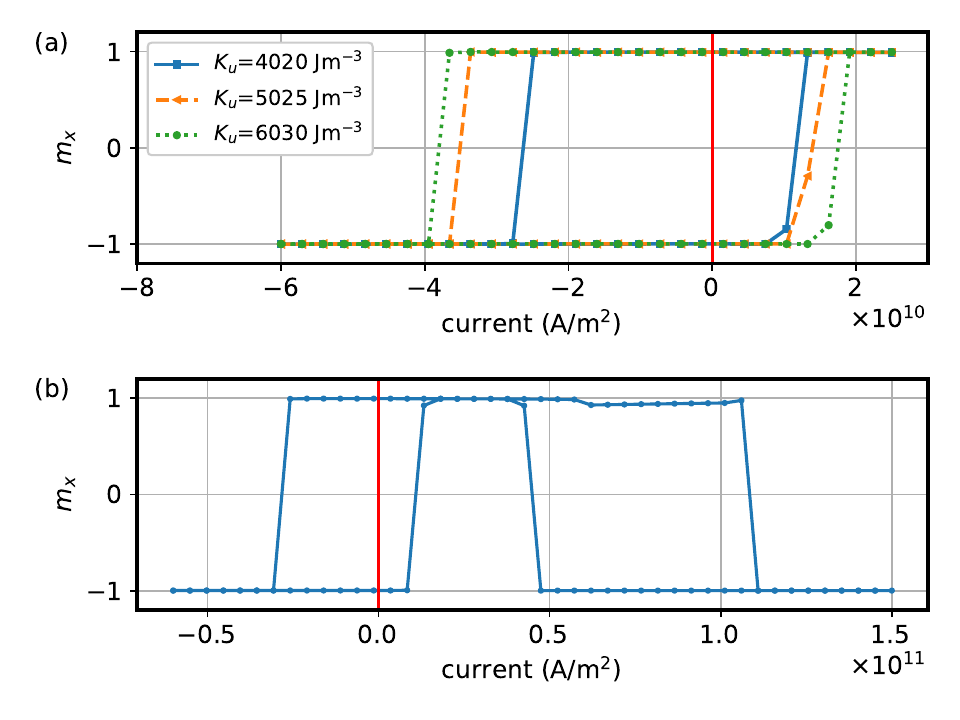}
\caption{(a) shows easy-axis hysteresis loops for the stack under investigation for different values of the anisotropy constants. (b) depicts the arise of a second loop due to the quadratic behavior of the dampinglike torque component for $K_u = 4020\text{ Jm}^{-3}$. In both figures, a vertical red line indicates $j=0\text{ Am}^{-2}$. The pulse duration for every data point is $30\text{ ns}$.}
\label{fig:hysteresis}
\end{figure}

Hysteresis loops of the same system are shown in figure~\ref{fig:hysteresis}. In contrast to the switching behavior simulations, the hysteresis loops are obtained by applying pulses of fixed current for a duration of $30\text{ ns}$ each. In sub-figure (a), the easy-axis hysteresis loop is shown for different values of the anisotropy constant $K_u$. As expected, an increase in the critical current for increasing values of $K_u$ is observed. Figure~\ref{fig:hysteresis} (b) shows a wider range of applied currents. Here, a second hysteresis loop can be seen for positive currents (positive voltage). This second loop arises from the sign-reversal of the dampinglike torque component, which was discussed earlier and is illustrated in figure~\ref{fig:voltage_dependence_tau} (b). \\
Experimental observation of a second loop on one voltage branch are presented by Devolder et al. in reference~\cite{Devolder_2020_PhysRevB} in the context of the well-known back-hopping phenomenon. Reference~\cite{Devolder_2020_PhysRevB} picks up on the sign-reversal of the dampinglike torque for large bias voltages in reference~\cite{Theodonis_2006_PhysRevLett} as possible explanation. However, the additional loop in reference~\cite{Devolder_2020_PhysRevB} is located after the ${\text{P }\rightarrow\text{ AP}}$ transition.\\ 
The back-hopping effect usually manifests itself in the coexistence of both the $\text{P}$ and $\text{AP}$ state for a single voltage value. The effect can be observed either on one voltage branch (e.g. in reference~\cite{Oh_2009_NatPhys}) or both (e.g. in reference~\cite{Min_2009_JApplPhys}). The idea that the back-hopping effect is closely related to the behavior of the dampinglike torque in MTJs should be taken with great caution and should be subjected to further research for the following reasons: Firstly, although it might lead to a second magnetization reversal event on the positive voltage branch, the quadratic behavior would not be able to explain back-hopping on both voltage-branches. Secondly, time-resolved studies of the back-hopping events carried out in reference~\cite{Devolder_2020_PhysRevB} in combination with the computational results from reference~\cite{Abert_2018_PhysRevAppl} indicate a coexistence of both a $\text{P}$ and $\text{AP}$ state for a single voltage and a cyclic process in which the spin-polarization-layer and the free-layer exert critical dampinglike torque on each other and thus leading to alternating switching of both layers.

\section{Conclusion}
We presented a highly performant algorithmic strategy that couples NEGF-based tunnel-current calculations with magnetization dynamic simulations in a micromagnetic context. This was achieved by utilizing the constant nature of both the fieldlike and dampinglike torque components with respect to the angle between reference-layer and free-layer for fixed voltages. This method can be easily extended to torque components with an angular dependence or fixed-current simulations by the introduction of a look-up table. The components can then be interpolated from the actual angle between the two magnetization directions in every time step of LLG integration. While more computationally expensive than it's fixed voltage counterpart, this strategy is still very performant, since it requires computationally expansive matrix inversions only during the setup phase.\\\\
The simulations carried out in this piece of work show ${\text{P}\rightarrow\text{AP}}$ switching times around $2-6\text{ ns}$ for voltages above $200\text{ mV}$. For ${\text{AP}\rightarrow\text{P}}$ switching, the shortest critical pulse duration is slightly above $7.5\text{ ns}$ at around $350\text{ mV}$. We have also reported the emergence of a second easy-axis hysteresis loop on the positive voltage branch due to the quadratic nature of the dampinglike torque in the fully-coherent NEGF regime. This somewhat unexpected feature should be further investigated as it might offer some further insight into the still not fully understood back-hopping in STT-MRAM cells.\\\\

\section*{Acknowledgments}
This research was enabled by the financial support of the FWF project I 4917.

\bibliographystyle{/home/peter/Uni/literature/bib/peter.bst}
\bibliography{bib/fun_amm.bib}

\end{document}